\documentclass[a4paper,fleqn]{cas-dc}

\usepackage[numbers]{natbib}

\begin{document}
\let\WriteBookmarks\relax
\def\floatpagepagefraction{1}
\def\textpagefraction{.001}
\shorttitle{}
\shortauthors{}

\title [mode = title]{$\alpha$-cluster formation in heavy $\alpha$-emitters within a multistep model}



\author[1]{J. M. Dong}
\address[1]{Institute of Modern Physics, Chinese Academy of Sciences, Lanzhou 730000, China}
\address[1]{School of Nuclear Science and Technology, University of Chinese Academy of Sciences, Beijing 100049, China}
\cormark[1]

\author[2]{Q. Zhao}
\address[2]{School of Nuclear Science and Technology, Lanzhou University, Lanzhou 730000, China}

\author[3]{L. J. Wang}
\address[3]{School of Physical Science and Technology, Southwest University, Chongqing 400715, China}

\author[1]{W. Zuo}

\author[4]{J. Z. Gu}
\address[4]{China Institute of Atomic Energy, P. O. Box 275(10), Beijing 102413, China}

\cortext[cor1]{dongjm07@impcas.ac.cn}

\begin{abstract}
$\alpha$-decay always has enormous impetuses to the development of
physics and chemistry, in particular due to its indispensable role
in the research of new elements. Although it has been observed in
laboratories for more than a century, it remains a difficult problem
to calculate accurately the formation probability $S_\alpha$
microscopically. To this end, we establish a new
model, i.e., multistep model, and the corresponding formation
probability $S_\alpha$ values of some typical $\alpha$-emitters are
calculated without adjustable parameters. The experimental
half-lives, in particular their irregular behavior around a shell
closure, are remarkably well reproduced by half-life laws combined
with these $S_\alpha$. In our strategy, the cluster formation is a
gradual process in heavy nuclei, different from the situation that
cluster pre-exists in light nuclei. The present study may pave the
way to a fully understanding of $\alpha$-decay from the perspective
of nuclear structure.
\end{abstract}




\maketitle


$\alpha$-decay is a typical radioactive phenomenon in which an
atomic nucleus emits a helium nucleus spontaneously. As one of the
most important decay modes for heavy and superheavy nuclei, it was
regarded as a quantum-tunneling effect firstly in the pioneering
works of Gamov, Condon and Gurney in 1928~\cite{Gamov,Condon}, which
provided an extremely significant evidence supporting the
probability interpretation of quantum mechanics in the early stage
of nuclear physics. However, a full understanding of $\alpha$-decay
mechanism and hence an accurate description of the half-life, have
not been settled yet. The critical problem lies in how to understand
the mechanism of $\alpha$-cluster formation and compute the
formation probability, known as a long-standing problem for nuclear
physics for more than eighty years that has attracted considerable
interest continuously~\cite{Review1998}. The $\alpha$-decay is
really understood only if the formation probability $S_\alpha$ can
be well determined microscopically.

The investigation of the $\alpha$-formation probability also
promotes the exploration of cluster structures in nuclei. Actually,
numerous experimental observations have already revealed clustering
phenomena in some light nuclei, such as the famous Hoyle state in
stellar nucleosynthesis that exhibits a structure composed of three
$\alpha$-particles~\cite{Hoyle}. The theoretical exploration of the
mechanism of cluster formation has been a hot topic in nuclear
physics~\cite{cluster1,cluster2,cluster3}. For heavy nuclei, a novel
manifestation of $\alpha$-clustering structure, namely, "$ \alpha +
^{208}$Pb" states in $^{212}$Po was revealed experimentally by their
enhanced $E1$ decays~\cite{E1}. Yet, it is still an open question
that whether or not the light and heavy nuclei share the same
mechanism of cluster formation.

Importantly, $\alpha$-decay has far-reaching implications in the
research of superheavy nuclei (SHN)~\cite{Triplet2,SHN}. Since an
"island of stability" of SHN was predicted in the 1960s,
experimental efforts worldwide have continuously embarked on such
hugely expensive programs since it is always at the exciting
forefront in both chemistry and physics~\cite{NAT2018,SHN}. However,
the exact locations of the corresponding nuclear magic numbers
remain unknown, and theoretical approaches to date do not yield
consistent predictions. The direct measurement of nuclear binding
energies and detailed spectroscopic studies of SHN with $Z > 110$
have been beyond experimental capabilities~\cite{NAT1,NAT2,NAT3},
therefore, to uncover their underlying structural information, one
has to resort to the mere knowledge about measured $\alpha$-decay
energies and half-lives~\cite{NAT3}. The formation probability, if
available microscopically, is of enormous importance to change this
embarrassing situation in combination with accurately measured
$\alpha$-decay properties.

Because of its fundamental importance, theoretically, the
exploration of the $\alpha$-particle formation can be traced back to
1960~\cite{Mang1960}, which triggered extensive investigations with
shell models~\cite{Review1998,Mang1976,Arima1979,NA2012}, Bardeen-Cooper-Schriffer (BCS)
models~\cite{Review1998,BCS1962,BCS1964} and Skyrme energy density
functionals~\cite{Ward2013} later. The formation amplitude is
regarded as the overlap between the configuration of a parent
nucleus and the one described by an $\alpha$-particle coupled to the
daughter nucleus. In particular, $^{212}$Po as a typical
$\alpha$-emitter with two protons and two neutrons outside the
doubly magic core $^{208}$Pb, was discussed extensively.
Nevertheless, these calculations disagree on the decay width, and
underestimate it substantially~\cite{Review1998,BCS1964,Shell}. To
improve the calculations, the shell model combined with a cluster
configuration, was proposed with a treatment of all correlations
between nucleons on the same footing~\cite{Varga}. Yet, these
calculations tend to be difficult to generalize for nuclei more
complex than $^{212}$Po. Over the past two decades, several new
approaches have been put forward to calculate the formation
probability $S_\alpha$ in different frameworks, including the
pairing approach~\cite{BCS1}, $N_nN_p$ scheme~\cite{NnNp},
quantum-mechanical fragmentation theory~\cite{FT}, cluster formation
models~\cite{CFM1,CFM2,CFM3}, a quartetting wave function
approach~\cite{Po212,QWFA1,QWFA2}, internal barrier penetrability
approach~\cite{BP}, statistical method~\cite{Dong20091}, some
empirical relations~\cite{DNP1,SEC1,SEC2,SEC3}, and extraction
combined with experimental
data~\cite{EXT1,EXT2,EXT3,EXT4,EXT5,EXT6}. Although great efforts
have been made and considerable progress has been achieved, no fully
satisfactory approach has yet been found until now.

To explore the $\alpha$-particle formation
probability $S_\alpha$, we propose a multistep model, and calculate
the $S_\alpha$ explicitly without introducing any adjustable
parameter with the help of a self-consistent energy density
functional theory.

\begin{figure*}
\begin{center}
\includegraphics[width=0.80\textwidth]{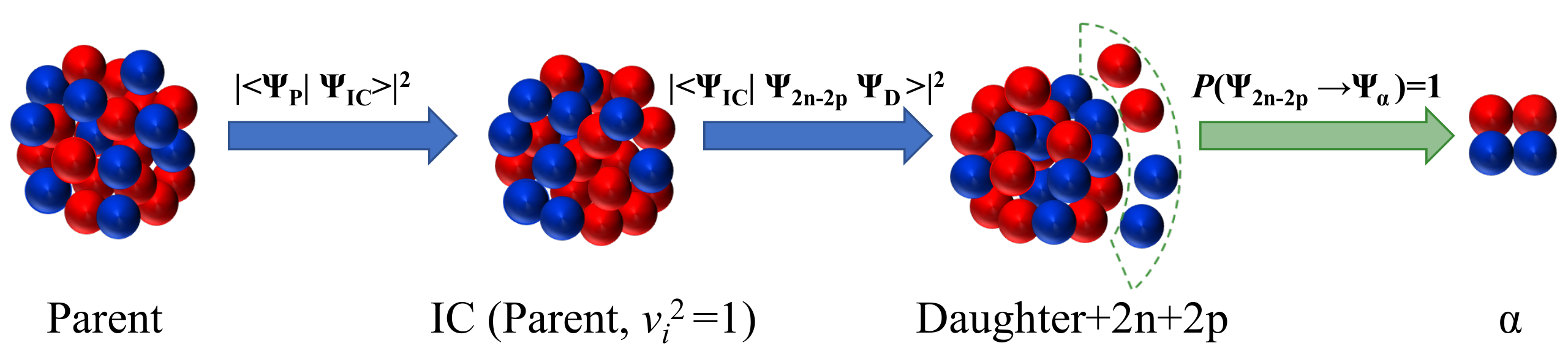}
\caption{(Color online) Schematic illustration of the physical
picture for an $\alpha$-cluster formation in the multistep model
through one of many pathways. The intermediate configuration (IC) is
the parent state but with a neutron level $i_n$ and a proton level
$i_p$ fully occupied ($v_i^2=1$). These four nucleons filling the
two single-particle levels will jump to the unoccupied $i_n$- and
$i_p$-levels of the daughter nucleus, and then automatically
assemble into an $\alpha$-particle finally. }\label{fig1}
\end{center}
\end{figure*}

Before we explore the $\alpha$-cluster formation
probability, we first discuss briefly the proton spectroscopic
factor of proton radioactivity. The initial state is proton
quasiparticle excitations of parent BCS
vacuum $\alpha _{i}^{\dagger }|$BCS$\rangle _{\text{P}}$ and the
final state is $c _{i}^{\dagger }|$BCS$\rangle _{\text{D}}$ with
$c_{i}^{\dagger }|$BCS$\rangle _{\text{D}}=\left[
u_{i}^{(\text{D})}\alpha
_{i}^{\dagger }+v_{i}^{(\text{D})}\alpha _{\overline{i}}\right] |$BCS$%
\rangle _{\text{D}}$~\cite{DSD2006}.
The spectroscopic factor (for spherical nuclei) is then given by $S_{p}=|_{\text{D}}\langle $BCS$|c_{i}\alpha _{i}^{\dagger }|$BCS$\rangle _{%
\text{P}}|^{2}\approx (u_{i}^{(\text{D})})^{2}$ \cite{DSD2006,Bonn}, where $(u_{i}^{(D)})^{2}=1-(v_{i}^{(D)})^{2}$ is the
probability that the spherical orbit of the emitted proton is empty
in the daughter nucleus. Accordingly, the $S_p$ can be iconically
interpreted as a probability that the blocked odd-proton in the $i_p$-orbit
of the parent nucleus jumps to the unoccupied $i_p$-orbit of the
daughter nucleus. With the inclusion of the
calculated $S_p$ from nuclear many-body approaches, the partial
half-lives for spherical proton emitters can be quite well
reproduced~\cite{Dong2009,Zhao2014}, indicating the success of the
strategy for proton spectroscopic factor.

Inspired by the proton radioactivity, our multistep
model is proposed, and a sketch is exhibited in Fig.~\ref{fig1} to
show schematically the physical picture of our model. However,
different from the proton radioactivity where the emitted proton
comes from the blocked odd-proton orbit in the parent nucleus, the
neutrons (protons) inside the emitted $\alpha$-particle could come
from any single-neutron (proton) level in principle. Therefore, we
introduce the intermediate configuration (IC) with mass number $A$
driven perhaps by quantum fluctuation and residual interactions, to
characterize which levels donate the complete four nucleons for the
$\alpha$-formation. The IC is a state that a single-neutron level
($i_n$) and a single-proton level ($i_p$) in the parent nucleus are
fully-occupied, namely, their occupation probabilities $v_i^2=1$ for
$i=i_n, i_p$ (which makes sure there are exactly four nucleons from
these two levels to generate an $\alpha$-particle), being analogous
to the blocked odd-proton in proton radioactivity. In fact, it is a
component of the initial parent state according to the
interpretation of quantum mechanics, for example,
the four-quasiparticle states in the picture of angular-momentum
projected models~\cite{PSM}. And there are many IC states and hence
many pathways to form the $\alpha$-particle, where Fig.~\ref{fig1}
just illustrates one of the pathways, and hence our strategy is
obviously distinguished from other models. These four quasiparticles
in the $i_n$- and $i_p$-levels are going to form an
$\alpha$-particle and the remaining $A-4$ nucleons accordingly form
a daughter nucleus, and the pathway via this IC is marked as
($i_n,i_p$) for the sake of the following discussion. Accordingly,
the probability to find a final configuration $\Psi _{\text{2n-2p}
}\Psi _{\text{D}}$ in the wavefunction of the parent nucleus $\Psi
_{\text{P}}$ through a given intermediate configuration $\Psi
_{\text{IC}}$ is
\begin{equation}
S_{\text{D} }^{(i_{n},i_{p})}=|\langle \Psi _{\text{P}}|\Psi _{\text{IC}%
}\rangle |^{2}\cdot |\langle \Psi _{\text{IC}}|\Psi _{\text{2n-2p}}\Psi _{%
\text{D}}\rangle |^{2}, \label{A}
\end{equation}
which is the formation probability of the daughter nucleus through
this pathway.

For a fully microscopic treatment, in principle,
many-body wave functions in the laboratory frame are need for
$|\Psi_\text{P} \rangle$, $|\Psi_\text{IC} \rangle$,
$|\Psi_\text{2n2p} \Psi_\text{D} \rangle$. While such treatments are
challenging and in this work a practical approximation is adopted,
i.e., we neglected beyond-mean-field effects and employ the
corresponding mean-field wave functions in the intrinsic frame. For
a deformed superfluid nucleus with nucleons paired by up and down
spins, within the BCS formulation, the overlap integral of $\langle
\Psi _{\text{P}}|\Psi _{\text{IC}}\rangle$ is written as
\begin{equation}
\langle \Psi _{\text{P}}|\Psi _{\text{IC}}\rangle =\underset{k}{\Pi }%
(u_{k}^{(\text{P})}u_{k}^{(\text{IC})}+v_{k}^{(\text{P})}v_{k}^{(\text{IC}%
)}),  \label{B}
\end{equation}
in which $v_{k}^{2}$ ($u_{k}^{2}$) represents the probability that
the two-fold degenerate $k$-th single-particle level is occupied
(unoccupied). The final state is written as $\Psi _{\text{2n-2p}
}\Psi _{\text{D}}=S_{i_{n}}^{\dagger }S_{i_{p}}^{\dagger }|$BCS$
\rangle _{D}$, where $S_{i_{n}}^{\dagger }$ ($S_{i_{p}}^{\dagger }$)
creates two neutrons (protons). As an
approximation, we assume the daughter even-even core can be
described within the BCS approach, just like that in proton
radioactivity~\cite{DSD2006}. Therefore, $\langle \Psi
_{\text{IC}}|\Psi _{\text{2n-2p} }\Psi _{\text{D}}\rangle $
expressed in terms of single-particle properties is given by
\begin{eqnarray}\label{C}
&&\langle \Psi _{\text{IC}}|\Psi _{\text{2n-2p}}\Psi
_{\text{D}}\rangle   \nonumber \\
&=& \left( \underset{i=i_{n},i_{p}}{\Pi }u_{i}^{(\text{D})}\right) ^{3}\underset%
{k\neq i}{\Pi }\left[ u_{k}^{(\text{D})}u_{k}^{(\text{IC})}+v_{k}^{(\text{D}%
)}v_{k}^{(\text{IC})}\langle \phi _{k}^{(\text{D})}|\phi _{k}^{(\text{IC}%
)}\rangle ^{2}\right],
\end{eqnarray}
where $\phi _{k}$ denotes the normalized single-particle
wavefunction. Iconically, the four nucleons
escaping from the IC jumps into the unoccupied $i_n$- and
$i_p$-levels of the daughter nucleus, with a probability
$(u_{i_{n}}^{(\text{D})}u_{i_{p}}^{(\text{D})})^{4}$.

We make the following two assumptions: 1) {\it The formation
probability of the $\alpha$-cluster is identical to that of the
daughter nucleus, i.e., $S_{\alpha }^{(i_{n},i_{p})}=S_{\text{D}
}^{(i_{n},i_{p})}$}, that is, the formation of the $\alpha$-particle
is achieved accordingly once the daughter nucleus is generated.
This means the four nucleons escaping from the IC
jumping into the unoccupied $i_n$- and $i_p$-levels of the daughter
nucleus are expected to automatically assemble into an
$\alpha$-particle at nuclear surface spontaneously. Namely, the
transition probability from $\Psi _{\text{2n-2p}}$ to an actual
$\alpha$-cluster state $\Psi _{\alpha}$ is $P(\Psi _{\text{2n-2p}}
\rightarrow \Psi _{\alpha})=1$ near the low-density nuclear surface.
It is generally believed that an $\alpha$-cluster is formed only in
regions that the nuclear matter density is low, and there is a
similar case that an $\alpha$-like state generates automatically
below $\rho ^{\text{Mott}}\simeq 0.03$ fm$^{-3}$ in quartetting wave
function approach~\cite{Po212,QWFA1,QWFA2}. Intuitively, that the
nucleons spontaneously organize into an doubly magic
$\alpha$-particle, makes the system more stable. 2) {\it Each
pathway for the formation process is expected to be independent of
the others}. We sum over all pathways (i.e., through different ICs)
to eventually achieve the $\alpha$-particle formation probability
via
\begin{eqnarray}
S_{\alpha } &=&\underset{(i_{n},i_{p})}{\sum
}S_{\text{D}}^{(i_{n},i_{p})}
\notag \\
&=&\underset{(i_{n},i_{p})}{\sum }\Bigg\{\underset{k^{\prime }}{\Pi
}\left(
u_{k^{\prime }}^{(\text{P})}u_{k^{\prime }}^{(\text{IC})}+v_{k^{\prime }}^{(%
\text{P})}v_{k^{\prime }}^{(\text{IC})}\right) ^{2}\cdot \underset{%
i=i_{n},i_{p}}{\Pi }\left( u_{i}^{(\text{D})}\right) ^{6}  \notag \\
&&\cdot \underset{k\neq i}{\Pi }\left[ u_{k}^{(\text{D})}u_{k}^{(\text{IC}%
)}+v_{k}^{(\text{D})}v_{k}^{(\text{IC})}\langle \phi
_{k}^{(\text{D})}|\phi _{k}^{(\text{IC})}\rangle ^{2}\right]
^{2}\Bigg\},  \label{AA}
\end{eqnarray}
with Eqs.~(\ref{A}-\ref{C}). The dimensionless formation probability
here is the expectation value of the $\alpha$-cluster number that
can be emitted. The stationary-state description of a time-dependent
cluster formation process is a quite good approximation and
simplifies the problem enormously~\cite{Review1998}, which is widely
used at present for $\alpha$-decay. It is valid because half-lives
of $\alpha$-emitters are very long ($10^{-6}-10^{17}$ s) compared
with the ``periods" of nuclear motion ($10^{-21}$ s) and hence in
the time evolution of a decaying state the nucleons has a large
number of opportunities to get clustered and to get the clusters
dissolved before it can actually escaped from the
nucleus~\cite{Review1998}.

Our approach involves the structure of both parent and daughter
nuclei, but does not involve an intrinsic state or a localized
density distribution of the $\alpha$-cluster, being significantly
different from the standard shell or BCS models where a Gauss-shaped
intrinsic $\alpha$-cluster wavefunction is introduced. The
quartetting wave function approach is a successful method proving a
reasonable behavior for the $S_\alpha$ of even-even Po
isotopes~\cite{QWFA1}, which does not involve such an intrinsic
$\alpha$-cluster wavefunction to calculate $S_\alpha$ either, and
does not employ the overlap of the wavefunctions between the initial
and final states. The wavefunction of the bound state for the
center-of-mass motion of four correlated nucleons is obtained by
solving the corresponding Schr\"odinger equation, and then the
$S_\alpha$ is calculated by integrating the modular square of this
wavefunction in the region below Mott density since an $\alpha$-like
state generates automatically at such low
densities~\cite{Po212,QWFA1,QWFA2}. Different substantially from
this quartetting wave function approach, the $S_\alpha$ in our work
is still based on the concept of overlap integrals, and finally can
be calculated with the compact expression of Eq. (\ref{AA}) with the
help of existing many-body approaches without introducing any
adjustable parameter.

The single particle properties in Eq. (\ref{AA}) are determined
within the framework of a covariant density functional (CDF)
approach starting from an interacting Lagrangian
density~\cite{RMF1,RMF2,RMF3,RMF4}. The nuclear CDF employed in
self-consistent calculations is parameterized by means of about ten
coupling constants that are calibrated to basic properties of
nuclear matter and finite nuclei, which enables one to perform an
accurate description of ground state properties and collective
excitations over the whole nuclear chart~\cite{RMF2,RMF3,RMF4}, and
has become a standard tool in low energy nuclear structure. The
explicit calculations are carried out based on a standard code
DIZ~\cite{Ring} for deformed nuclei, with the NL3
interaction~\cite{GLJK} for the mean-field and the calibrated D1S
Gogny force~\cite{Wang1} for the pairing channel. The NL3 parameter
set has been used with enormous success in the description of a
variety of ground-state properties of spherical, deformed and exotic
nuclei~\cite{GLJK,BGT}, and the calibrated D1S Gogny force enables
one to well reproduce the odd-even staggerings on nuclear binding
energies~\cite{Wang1}. We concentrate on the even-even Po, Rn and Ra
isotopes with spherical or near-spherical shapes, because their
$\alpha$-decays tend to have large branching ratios (100\% in most
cases) and their corresponding half-lives were best measured
experimentally~\cite{Mass2012}. On the other hand, these
$\alpha$-decay cases usually do not involve excited states and
angular momentum transfers, and thus serve as an optimal testing
ground to examine our
model. Moreover, the values of overlap integrals $\underset{k}{\Pi }\langle \phi _{k}^{(\text{D})}|\phi _{k}^{(\text{%
IC})}\rangle $ for these nuclei can be taken as unity. The products
in Eqs. (\ref{B},\ref{C}) along with the summation in Eq. (\ref{AA})
are truncated at 5 MeV for the single-nucleon spectra to achieve
convergence.

\begin{table}
\centering
\caption{The rms deviations $\sqrt{\langle \sigma ^{2}\rangle }$ and average deviations $\langle
\sigma \rangle $ for the VSF and UDL with $S_\alpha=1$ and $S_\alpha \neq 1$.}
\label{Table1}
\begin{tabular}{ccccccc}
\hline
Formulas  & $\sqrt{\langle \sigma ^{2}\rangle }$  & $\langle \sigma
\rangle
$ \\
\hline
VSF ($S_\alpha=1$) & 0.360  & 0.296   \\
VSF ($S_\alpha \neq 1$)& 0.109 & 0.0861   \\
UDL ($S_\alpha=1$)& 0.316  & 0.268 \\
UDL ($S_\alpha \neq 1$)& 0.0888  & 0.0812 \\
\hline
\end{tabular}
\end{table}

To assess the validity of our multistep model, we
explore the role of the formation probability $S_\alpha$ in
half-life calculations. The widely accepted formulas, i.e., the
semi-empirical Viola-Seaborg formula (VSF)~\cite{VSF} and the
universal decay law (UDL) based on the $R$-matrix
expression~\cite{UDL} are employed, which are respectively given as
\begin{eqnarray}
\log _{10}T_{1/2} &=&\frac{aZ+b}{\sqrt{Q_{\alpha }}}+cZ+d-\log
_{10}S_{\alpha }, \label{D} \\
\log _{10}T_{1/2} &=&a\chi ^{\prime }+b\rho ^{\prime }+c-\log
_{10}S_{\alpha } \label{E},
\end{eqnarray}
with \begin{eqnarray} \chi ^{\prime } &=&2(Z-2)\sqrt{\frac{A_{\alpha
d}}{Q_{\alpha }}},A_{\alpha
d}=4(A-4)/A, \nonumber\\
\rho ^{\prime } &=&\sqrt{2A_{\alpha d}(Z-2)\left[
(A-4)^{1/3}+4^{1/3}\right] }.\nonumber
\end{eqnarray}
$Z$ ($A$) is the proton (mass) number of a given parent nucleus. The
decay energy $Q_\alpha$ and half-life $T_{1/2}$ are in units of MeV
and second, respectively. $S_\alpha=1$ ($S_\alpha \ne 1$)
corresponds to the results without (with) the inclusion of the
formation probability.

\begin{figure}
\begin{center}
\includegraphics[width=0.45\textwidth]{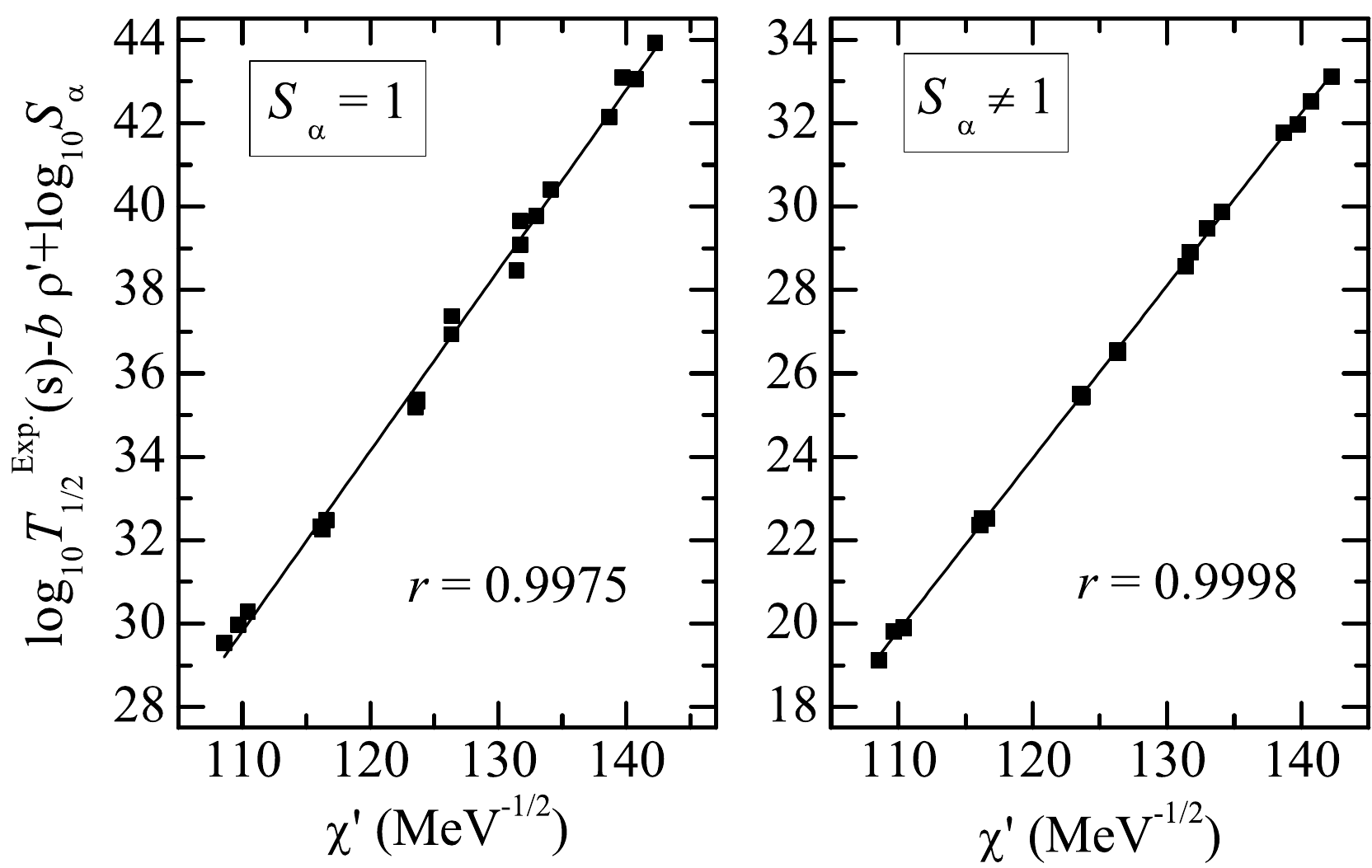}
\caption{$\log _{10}T_{1/2}^{\text{Exp.}}-b\rho ^{\prime }+\log
_{10}S_{\alpha }$ as a function of $\chi ^{\prime }$ obtained with
the UDL for $S_\alpha=1$ (left) and $S_\alpha \neq 1$ (right). The
coefficient $b$ is fixed at the fitted value in the two cases,
respectively. The straight lines are given by $a \chi ^{\prime } +
c$. Here $r$ is the corresponding correlation
coefficient.}\label{fig2}
\end{center}
\end{figure}

The fitting procedures are performed in the cases of $S_\alpha=1$
and $S_\alpha \neq 1$ respectively to test whether or not the
predicted $S_\alpha$ could improve substantially the accuracy of the
two formulas. It is worth pointing out that the UDL has already
included the logarithmic formation amplitude which is assumed to be
linearly dependent upon $\rho ^{\prime }$. Therefore, in Eq.
(\ref{E}), the $\rho ^{\prime }$-dependent formation probability is
replaced by the presently calculated $S_\alpha$. The
root-mean-square (rms) deviations $\sqrt{\langle \sigma ^{2}\rangle
}$ and average deviations $\langle \sigma \rangle $ for the two
formulas with $S_\alpha=1$ and $S_\alpha \neq 1$ are summarized in
Table~\ref{Table1}. The inclusion of the $S_\alpha$ indeed greatly
improves the accuracy of both the VSF and UDL. The UDL with a solid
physical ground but less parameters, works better than the VSF, and
reproduces the available experimental half-lives within a factor of
2 in the case of $S_\alpha=1$. Yet, when the microscopically
calculated $S_\alpha$ is included, the deviation of the refitting is
reduced down to around $20\%$. The good agreement between the
calculated half-lives and the experimental data is quite
encouraging, indicating the reliability of the formation probability
$S_\alpha$ given by the multistep model. In Fig.~\ref{fig2}, we plot
the UDL fittings but replace the half-lives with the experimental
values to more visually reveal the role of $S_\alpha$, and that the
inclusion of the $S_\alpha$ systematically improves the agreement
with data is exhibited. The highly linear correlation is displayed
for $S_\alpha \neq 1$, with a correlation coefficient as high as
$r=0.9998$, suggesting the success of our formation probability and
the validity of the two assumptions.

\begin{figure}
\begin{center}
\includegraphics[width=0.45\textwidth]{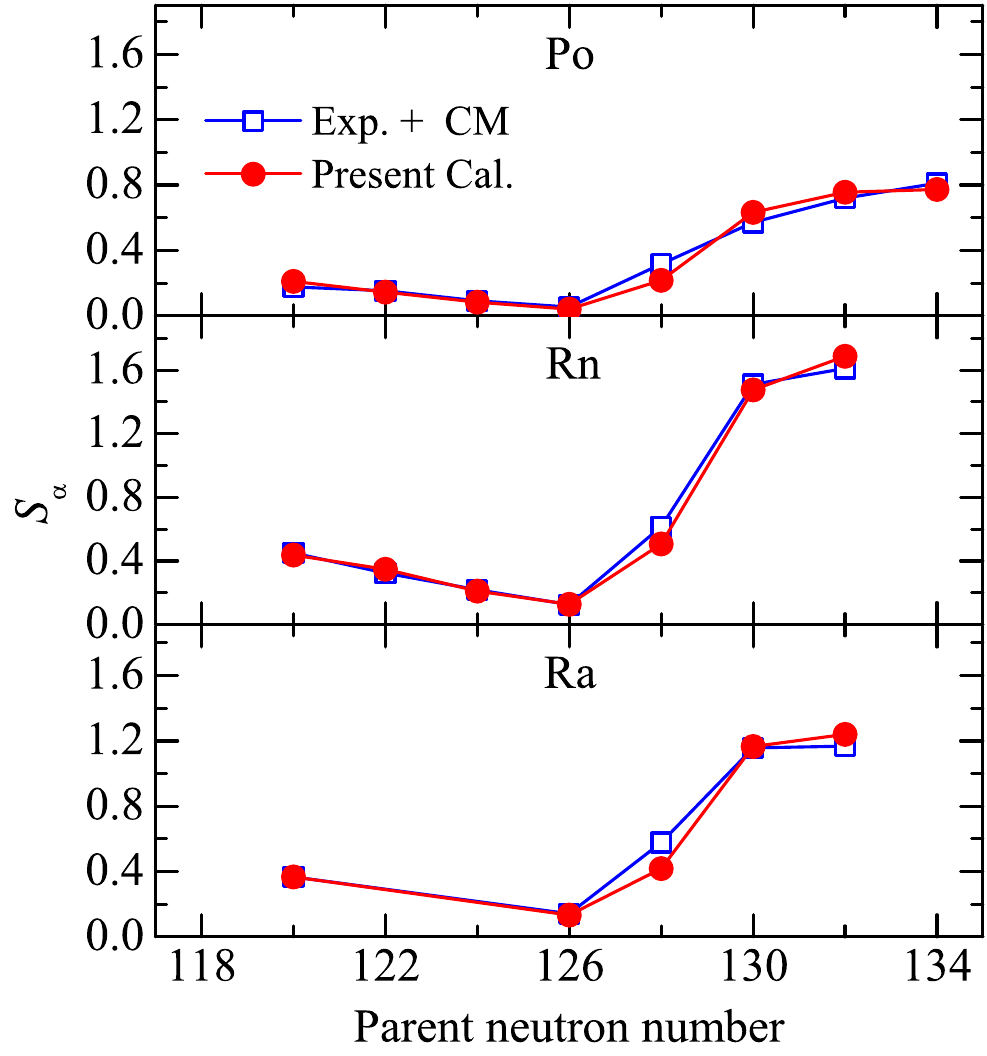}
\caption{(Color online) Calculated formation probability $S_\alpha$
values within the multistep model for the Po, Rn and Ra isotopes,
compared with those extracted by using the experimental half-lives
in combination with the CM calculated penetration
probabilities.}\label{fig3}
\end{center}
\end{figure}

Furthermore, $S_\alpha$ is extracted in turn by using the ratio of
the theoretical half-life to the experimentally observed value. The
barrier penetrability of $\alpha$-particle, is achieved
theoretically by the WKB approximation which turns out to work
excellently~\cite{Dong2011}, where the potential barrier is
constructed by a simple "Cosh" potential plus the Coulomb barrier
model (CM)~\cite{CM}. Here the extracted $S_\alpha$ should be
considered as a relative value. By selecting an optimal constant
assault frequency, the extracted $S_\alpha$ values with varying
neutron number $N$ are compared with the results given by the
multistep model in Fig.~\ref{fig3}. In sharp contrast with
half-lives, the $S_\alpha$ values are located in a relatively narrow
range, leading to the success of the empirical half-life laws even
when $S_\alpha$ is not included. The $S_\alpha$ values follow the
similar behavior with regard to the Po, Rn and Ra isotopic
chains--that is, gradually drop with increasing neutron number up to
the spherical magic number $N=126$, attributed to the increased
stability of isotopes when approaching the magic number, and then
they increase drastically with neutron number. Such a general trend
of the extracted $S_\alpha$ is successfully reproduced within our
method. Typically, the $S_\alpha$ of $^{212}$Po, being expected to
be large owing to its two protons and two neutrons outside the shell
closure core $^{208}$Pb, is about six times larger than that of its
neighbor $^{210}$Po. The weight of the cluster component in
$^{212}$Po is large, which is exactly what one needs to
simultaneously describe the B(E2)~\cite{BE2-2012} and the absolute
$\alpha$-decay width~\cite{Review1998,Varga} within the shell model
plus a cluster component. The distinct behavior that $S_\alpha$
varies abruptly when the magic number is crossed, confirms that the
particularly significant role of shell effects on $S_\alpha$ is
reasonably accounted for in Eq. (\ref{AA}) via the single-particle
properties. As one expects, $S_\alpha$ reaches its minimum at the
shell closure $N=126$ as the result of the well-known shell
stability that strongly enhances the nuclear binding.

\begin{figure}
\begin{center}
\includegraphics[width=0.5\textwidth]{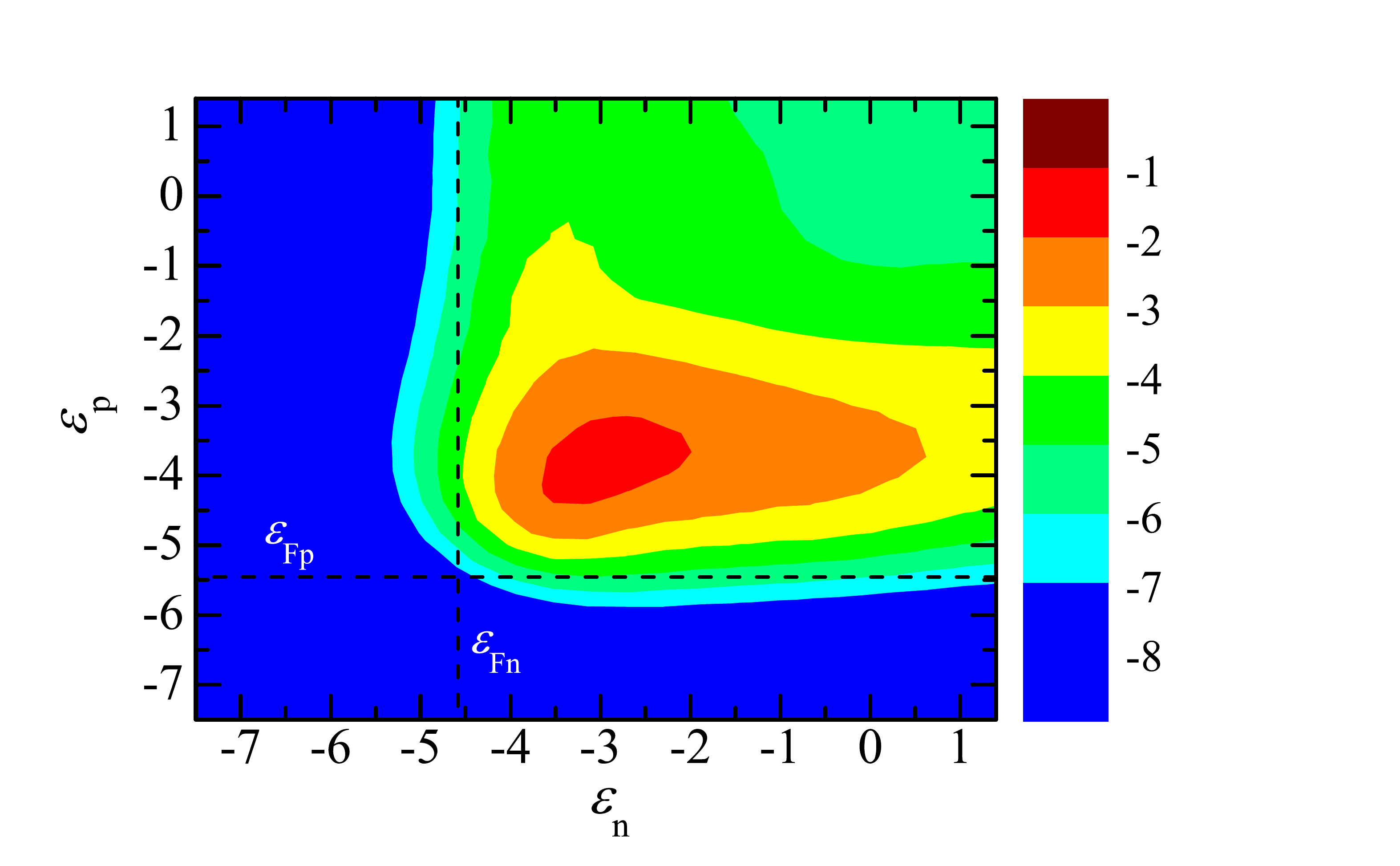}
\caption{Contour plot of $\log_{10}S_{\alpha }^{(i_{n},i_{p})}$
versus the single-particle energies (in units of MeV) of the $i_n$-
and the $i_p$-levels for the illustrative example of $^{212}$Po to
show the role of each pathway. The dashed lines denote the
corresponding Fermi energies $\varepsilon _{\text{Fn}}$ and
$\varepsilon _{\text{Fp}}$ for neutrons and protons, respectively.
  }\label{fig4}
\end{center}
\end{figure}

In order to analyze the contribution of each pathway in Eq.
(\ref{AA}) to $S_\alpha$, Fig.~\ref{fig4} illustrates $S_{\alpha }^{(i_{n},i_{p})}=|\langle \Psi _{\text{IC}}|\Psi _{\text{P}%
}\rangle |^{2}\cdot |\langle \Psi _{\text{2n-2p} }\Psi _{\text{D}}|\Psi _{\text{IC}%
}\rangle |^{2}$ versus the single neutron and single proton energies
($\varepsilon _{\text{n}}$, $\varepsilon _{\text{p}}$) by taking the
typical nucleus $^{212}$Po as an example. The formation probability
in Eq.~(\ref{AA}) is predominantly determined by the pathways
belonging, in the IC, to the fully-occupied neutron and proton
levels $i_n$ and $i_p$ slightly above the Fermi surfaces, i.e.,
these levels are the major nucleon donors to constitute the emitted
$\alpha$-particle. The contributions from other pathways drop
sharply when the fully-occupied levels ($i_n,i_p$) gradually go away
from these leading ones.

It is well-known that the concept of $\alpha$-clustering is
essential for understanding the structure of light nuclei. In some
cases, light nuclei behave like molecules composed of clusters of
protons and neutrons, such as the definite $2\alpha$ cluster
structure of $^8$Be~\cite{Twoalpha} reflected in a localized density
distributions. But whether such type of cluster structure exists or
not in heavy nuclei is uncertain. In our strategy, however, a
localized $\alpha$-cluster does not pre-exist inside a parent
nucleus, but is generated during the decay process in our model.
Therefore, the scenario that a continuous formation and breaking of
the $\alpha$-cluster until it escapes randomly from the parent
nucleus~\cite{Review1998}, is supported. As a result, the mechanism
of $\alpha$ formation in heavy $\alpha$-emitters is different
markedly from that in light nuclei.

Since the formation probability $S_\alpha$ is highly relevant to the
quantum-mechanical shell effects, the extracted $S_\alpha$ with a
high-precision UDL combined with experimentally measured
$\alpha$-decay properties, is of great importance for digging up
valuable structural information of SHN and exotic nuclei. The
correlations between the $Q_{\alpha}$ values of SHN have suggested
that the heaviest isotopes reported in Dubna do not lie in a region
of rapidly changing shapes~\cite{Dong2011-1}. Therefore, the
$S_\alpha$ versus proton number exhibits a behavior analogous to
isotopic chains shown in Fig.~\ref{fig3}, attributed to the nearby
shell closure. By employing the NL3 interaction, the heaviest
nucleus $^{294}\text{Og}$ ($Z=118$) and its isotonic neighbor
$^{292}\text{Lv}$ ($Z=116$), are predicted to be nearly spherical.
The calculated $S_\alpha$ is 0.66 for $^{292}\text{Lv}$ while it
reduces to 0.45 for $^{294}\text{Og}$ with the ratio of $S_\alpha
(^{292}\text{Lv})/S_\alpha (^{294}\text{Og})=1.5$, being indeed
similar to the trend shown in Fig.~\ref{fig3}, which is consistent
with the fact that the NL3 interaction itself predicts the adjacent
$Z=120$ as a magic number. On the other hand, with the UDL of Eq.
(\ref{E}) along with experimental data~\cite{SHN}, the extracted
ratio of $S_\alpha (^{292}\text{Lv})/S_\alpha (^{294}\text{Og})$ is
as high as $3.2^{+4.3}_{-1.8}$ ($4.5^{+5.8}_{-2.2}$) with half-life
data of $^{292}\text{Lv}$ from Ref.~\cite{Dub-116}
(Ref.~\cite{GSI-116}) where the significant uncertainties are due to
the low-statistics data for the measurements, and agrees marginally
with the above theoretical value. Hence the probable magic nature of
$Z=120$ is suggested. The future measurements with a much higher
accuracy for these nuclei together with their isotopes are
encouraged, which would pin down the proton magic number eventually.

Intriguingly, the superallowed $\alpha$-decay to doubly magic
$^{100}\text{Sn}$ was observed recently, which indicates a much
larger $\alpha$-formation probability than $^{212}\text{Po}$
counterpart~\cite{Sn100}. Within the framework of
the quartetting wave function approach, an enhanced $\alpha$-cluster
formation probability for $^{104}\text{Te}$ was found because the
bound state wavefunction of the four nucleons has a large component
at the nuclear surface~\cite{QWFA2}. Yet, a large $S_\alpha$ for
$^{104}\text{Te}$ is inconsistent with our prediction of
$S_\alpha=0.24$, suggesting the onset of an unusual type of nuclear
superfluidity for self-conjugate nuclei, i.e., the proton-neutron
pairing which is not well-confirmed at present. This isoscalar
pairing would considerably impact on the single-particle spectra and
hence the $S_\alpha$, and its absence in our CDF calculations leads
to the underestimated $S_\alpha$. Therefore, as a new way
independent of Ref.~\cite{NZP}, combined with precise $\alpha$-decay
measurements for $N \simeq Z$ nuclei, our approach for the
$S_\alpha$ could enable us to clarify this abnormal pairing
interaction in turn by employing CDF approaches with the inclusion
of a tentative isoscalar superfluidity.

In general, computing the formation probability $S_\alpha$ defined
in Eq. (\ref{AA}) by nuclear density functionals with a very high
accuracy, is out of reach at present because of the well-known fact
that the single-particle levels are not well-defined in the concept
of the mean-field approximation especially for well-deformed nuclei.
The approximate single-particle spectrum is perhaps
the primary cause which leads to the deviation of the theoretical
calculations. Nevertheless, based on two intuitive assumptions and
without any phenomenological adjustment, our strategy opens a new
perspective to account for the formation mechanism. This is highly
important to help one to uncover the underlying knowledge about
superheavy nuclei and isoscalar pairing, in combination with
experimental observations. For example, the experimentally measured
$\alpha$-decay properties of heaviest nuclei combined with our
calculated $S_\alpha$ suggest the probable proton-magic nature of
$Z=120$. To treat the many-body wave functions in a
laboratory frame beyond mean-field approximation, is planed as a
future work.


J. M. Dong thanks B. Zhou, W. Scheid, and C. Qi for helpful comments
and suggestions. This work is supported by the Strategic Priority Research Program of
Chinese Academy of Sciences (Grant No. XDB34000000), by the National Natural
Science Foundation of China (Grants Nos. 11775276, 11905175, 11975282, and 11675265), by the National
Key Program for S\&T Research and Development (Grant No.
2016YFA0400502), by the Youth Innovation Promotion Association of
Chinese Academy of Sciences under Grant
No. Y201871, by the Continuous Basic Scientific
Research Project (Grant No. WDJC-2019-13), by the Leading Innovation Project (Grant No. LC 192209000701), and
by the Continuous Basic Scientific Research Project (Grant No. WDJC-2019-13).



\end{document}